\documentstyle[prd,preprint,aps]{revtex}

\begin{document}
\tightenlines
%\draft
\parskip 0.3cm
\begin{titlepage}

\begin{centering}
{\large \bf Signatures for Majorana neutrinos in $e^- \gamma$
collider}\\ \vspace{.9cm}
{\bf J.Peressutti and O.A.Sampayo}\\
\vspace{.05in}
{\it  Departamento de F\'{\i}sica,
Universidad Nacional de Mar del Plata \\
Funes 3350, (7600) Mar del Plata, Argentina} \\ \vspace{.4cm}
 \vspace{.05in}
{\bf  Jorge Isidro Aranda}
\vspace{.05in}
\\ {\it Escuela de Ciencias F\'isico-Matem\'aticas,\\
Universidad Michoacana de San Nicolas de Hidalgo,\\ Morelia,
Michoacan, Mexico.}\\ \vspace{1.5cm}
\vspace{.08in}
{ \bf Abstract}
\\
\bigskip
\end{centering}
{\small
 We study the possibilities to detect Majorana neutrinos in
 $e^- \gamma$ colliders for different center of mass energies.
 We study the $W^- W^- l_j^{+}$($l_j^+\equiv e^+ ,\mu^+ ,\tau^+)$
 final state which are, due to leptonic
 number violation, a clear signature for intermediate Majorana
 neutrino contribution. Such a signal (final lepton have the opposite
 charge of the initial lepton) is not possible if
 the heavy
 neutrinos are Dirac particles. In our calculation we use the helicity
 formalism to obtain analytic expressions for the amplitude and we
 have considered that the intermediate neutrinos can be either on shell or
 off shell. Finally we present our results for the total
 cross-section and for the angular distribution of the final
 lepton. We also include a discussion on the expected
  events number as a function of the input parameters.}


\vspace{0.2in}
\vfill

%\begin{flushleft}
%MDP-HEP/98-01\\
%October 1998 \hfill
%\draft
%\end{flushleft}
\end{titlepage}
\vfill\eject

\section{\bf Introduction}

Massive neutrinos can come in two different types: as Dirac or
Majorana particles. Dirac fermions have distinct particles and
antiparticle degrees of freedom while Majorana fermions make no
such distinction and have half as many degrees of freedom. In this
conditions fermions with conserved charges (color, electric
charged, lepton number,...) must be of Dirac type, while fermions
without conserved charges may be of either type. If the neutrino
mass vanishes then both types are equivalent to two-component Weyl
fermions and the distinction between Dirac and Majorana neutrinos
vanishes \cite{neutmass} . New neutrinos could have large masses
and be of either type. If there are heavy neutrinos, then the
present and future experiments offer the possibility of
establishing their nature. The production of Majorana neutrinos
via $e^+e^-$ and hadronic collision have been extensively
investigated in the past
\cite{ma,datta,gluza,hoefer,cevetic,almeida}. In this work we
study the possibility of the $\gamma e^-$ linear collider to
produce clear signatures for Majorana neutrinos. The photon linear
collider \cite{ohgaki} may be the best alternative to the electron
positron colliders and furthermore, in the case of Photon linear
collider, we can control the initial photon polarization by the
inverse Compton scattering of the polarized laser by the electron
beam at NLC. Using these polarized high energy photon beam we have
the possibility to study in detail the interaction of Majorana
neutrinos and reject possible background. In this paper we discuss
the signatures for Majorana neutrinos in the reactions $\gamma e^-
\rightarrow W^- W^- l_j^{+}$($l_j^+\equiv e^+ ,\mu^+ ,\tau^+)$.
For the cross section calculation we have used the helicity
formalism. The phase space integration was done taking account
that the intermediate neutrinos can be either on shell or off
shell. Due the large CM energies of these colliders we have
considered that the mass of the final lepton vanishes. Moreover we
study distributions of the final leptons for different
polarizations of the initial photon.

For the couplings of the Majorana neutrinos we follow Ref
\cite{cevetic} starting with rather general lagrangian densities
for the interaction of $N$ with $W$ and light leptons $l_i$
($e,\mu,\tau$):
\begin{equation}
{\cal L}_{NWl}=-\sum_{j=1}^3 \frac{g B_L^{(j)}}{\sqrt{2}} \bar l_j
\gamma^{\mu} P_L N W^-_{\mu} + h.c.
\end{equation}
The heavy Majorana neutrino couples to the three flavors lepton
with couplings proportional to $B_L^{j}$, where $j$ labels the
family. The constant $g$ is the standard $SU(2)_L$ gauge coupling.

This $B_L^{j}$ parameters will affect the final results via the
combinations $H_1=\mid B_L^{(1)} \mid^2$ and $H=\sum_{j=1}^3\mid
B_L^{(j)} \mid^2$ in the following way: The cross-section is
proportional to $H_1$. This proportionally came from the
interaction of the initial electron with either of intermediary or
final $W$ boson (Fig.1). In the other hand the final lepton can be
either of $e^+$, $\mu^+$ or $\tau^+$ because this is allowed by
the interaction lagrangian (eq.1). All these possible final states
are a clear signal for intermediary Majorana neutrino and then we
sum the cross section over the flavors of the final lepton. This
sum produces a $H$ factor in the numerator of the total cross
section. In the other hand this cross-section also depends on $H$
through the total width $\Gamma_{N\rightarrow all}$ (eq.10) in the
Majorana neutrino propagator.

 In this work we have considered the complete
set of Feynman diagrams (Fig.1) that contribute at tree level to
$e^- \gamma\rightarrow W^-W^-l^+_j$ ($\rightarrow jets + l_j^+$)
with the light leptons $l_1^+=e^+,l_2^+=\mu^+,l_3^+=\tau^+$.

%%%%%%%%%%%%%%%%%%%%%%%%%%%%%%%%%%%%%%%%%%%%%%%%%%%%%%%%%%%%%%%%%%%%%%%%%
\section{Helicity Amplitude}

When the number of Feynman diagrams is increased, the calculation
of the amplitude is a rather unpleasant task. Some algebraic forms
can be used in it to avoid manual calculation, but sometimes the
lengthy printed output from the computer is overwhelming, and one
can hardly find the required results from it. The CALKUL
collaboration \cite{Causmaecker} suggested the Helicity Amplitude
Method
 (HAM) which can
simplify the calculation remarkably and hence make the manual calculation
realistic.

In this section we discuss the evaluation of the amplitudes at the
tree level for $\gamma e^{-}\rightarrow W^- W^-
l_j^{+}$($l_j^+\equiv e^+ ,\mu^+ ,\tau^+$)
 using the HAM. This method is a powerful technique for computing helicity
amplitudes for multiparticle processes involving massless spin-1/2
and spin-1 particles. Generalization of this method which
incorporates massive spin-1/2 and spin-1 particles, are given in
Ref. \cite{stirling}. This algebra is easy to program and more
efficient than computing the Dirac algebra.

The  Feynman diagrams, which contribute at the tree-level are
depicted in Fig.1 and the corresponding amplitudes can be
organized as follow

\begin{eqnarray}
i{\cal M}_1(\lambda)&=& i C
P_N(k_3+p_2)P_e(p_1+k_1)T_1(\lambda)+(k_2 \leftrightarrow k_3)
,\nonumber\\ i{\cal M}_2(\lambda)&=&-i C
P_N(k_2-p_1)P_e(p_2-k_1)T_2(\lambda)+(k_2 \leftrightarrow k_3),
\nonumber\\ i{\cal M}_3(\lambda)&=&-i C
P_W(k_3-k_1)P_N(k_2+p_2)T_3(\lambda)+(k_2 \leftrightarrow k_3) ,\\
i{\cal M}_4(\lambda)&=&-i C
P_W(k_3-k_1)P_N(k_2-p_1)T_4(\lambda)+(k_2 \leftrightarrow k_3)
,\nonumber
\end{eqnarray}

where $\lambda$ is the polarization of the photon, $C=M_N
\Lambda_M e g^2 B_L^1 B_L^j /2$ ($\Lambda_M$ is the phase factor
in the Fourier decomposition of the Majorana field $N(x)$;
$|\Lambda_M|^2=1$  \cite{neutmass} ) and $p_1$, $k_1$, $p_2$,
$k_2$ and $k_3$ are the 4-impulse of the particles $e^-$,
$\gamma$, $l^+$, $W^-$ and $W^-$ respectively. The corresponding
propagators are

\begin{eqnarray}
P_N(k)&=&\frac{(k^2-M_N^2)+i M_N \Gamma_N} {(k^2-M_N^2)^2+(M_N
\Gamma_N)^2},\nonumber\\
 P_W(k)&=&\frac{1}{k^2-m_W^2},\\
P_f(k)&=&\frac{1}{k^2-m_f^2},\nonumber
\end{eqnarray}
Following the Feynman rules \cite{neutmass} is straightforward to
obtain the $T$ amplitudes:
\begin{eqnarray}
T_1 &=& \bar v(p_1) \gamma_{\mu} (p\llap{/}_1+k\llap{/}_1)
\gamma_{\nu} \gamma_{\alpha} P_R v(p_2) \epsilon^{\mu}(k_1)
\epsilon^{\nu}(k_2) \epsilon^{\alpha}(k_3)
 \nonumber \\ T_2 &=&
\bar v(p_1) \gamma_{\mu} \gamma_{\nu}(p\llap{/}_2-k\llap{/}_1)
\gamma_{\alpha} P_R v(p_2) \epsilon^{\mu}(k_2) \epsilon^{\nu}(k_3)
\epsilon^{\alpha}(k_1) \nonumber \\ T_3 &=& \bar v(p_1)
\gamma_{\mu} \gamma_{\nu} P_R v(p_2) \left[\hspace{1mm}g^{\mu
\beta} - (k1-k3)^{\mu}(k1-k3)^{\beta}/m_W^2\right]  \\
&&\left[\hspace{1mm}g_{\rho \beta} (2 k_3-k_1)_{\lambda}+g_{\beta
\lambda} (2 k_1-k_3)_{\rho}-g_{\lambda \rho}
(k_1+k_3)_{\beta}\right] \epsilon^{\lambda}(k_1)
\epsilon^{\nu}(k_2) \epsilon^{\rho}(k_3) \nonumber \\
 T_4 &=& \bar v(p_1) \gamma_{\nu} \gamma_{\mu} P_R v(p_2)
\left[\hspace{1mm}g^{\mu \beta} -
(k1-k3)^{\mu}(k1-k3)^{\beta}/m_W^2\right]  \nonumber \\
&&\left[\hspace{1mm}g_{\rho \beta} (2 k_3-k_1)_{\lambda}+g_{\beta
\lambda} (2 k_1-k_3)_{\rho}-g_{\lambda \rho}
(k_1+k_3)_{\beta}\right] \epsilon^{\lambda}(k_1)
\epsilon^{\nu}(k_2) \epsilon^{\rho}(k_3) \nonumber
\end{eqnarray}

 In order to calculate these amplitudes we follow the
rules from helicity formalism and use identities of the type

\begin{equation}
\{\bar u_{\lambda}(p_{1})\gamma ^{\mu}u_{\lambda}(p_{2})\}\gamma_{\mu}=2u_{\lambda}(p_{2})\bar u_{\lambda}(p_{1})+2u_{-\lambda}(p_{1})\bar u_{-\lambda}(p_{2}),
\end{equation}

\noindent which is in fact the so called Chisholm identity, and

\begin{equation}
p\llap{/}=u_{\lambda}(p)\bar u_{\lambda}(p)+u_{-\lambda}(p)\bar u_{-\lambda}(p),
\end{equation}

\noindent defined as a sum of the two projections $u_{\lambda}(p)\bar u_{\lambda}(p)$
and $u_{-\lambda}(p)\bar u_{-\lambda}(p)$.

The spinor products are given by

\begin{eqnarray}
s(p_{i}, p_{j})&\equiv&\bar u_{+}(p_{i})u_{-}(p_{j})=-s(p_{j}, p_{i}),\nonumber\\
t(p_{i}, p_{j})&\equiv&\bar u_{-}(p_{i})u_{+}(p_{j})=[s(p_{j}, p_{i})]^{*}.
\end{eqnarray}

Using the above rules, which are proved in Ref. \cite{stirling}, we
can reduce many amplitudes to expressions involving only spinor products.

In order to add up the polarization of the $W$ vector bosons in
the final state we define two auxiliary lightlike 4-vectors for
each $W$ such that $k_i=r^1_i+r^2_i$, $(r^1_i)^2=(r^2_i)^2=0$ and
$(k_i)^2=m_W^2$ ($i=2,3$). We also introduce the object
$a_i^{\mu}=\bar u_{-}(r^1_i)\gamma^{\mu}u_{-}(r^2_i)$. As was
shown in Ref \cite{stirling} we will arrive at the correct result
for the cross section if we make the following replacements for
the outgoing W:

\begin{eqnarray}
 \epsilon^{\mu} &\rightarrow & a^{\mu} , \nonumber \\
\sum_{pol} \epsilon^{\mu} \epsilon^{*\nu} & \rightarrow &
\frac{3}{8\pi m_W^2} \int d\Omega a^{\mu}a^{*\nu}
\end{eqnarray}

In order to obtain the cross section we have to perform additional
two-dimensional integral but no accuracy will be lost since the
accuracy of Monte Carlo integration does not depend on the
dimensionality.

For the polarization of the initial photon we take \cite{stirling}
$\epsilon^{\mu}_{\lambda}(k)=N \bar
u_{\lambda}(k)\gamma^{\mu}u_{\lambda}(p)$ where $p^{\mu}$ is any
lightlike vector not collinear to $k^{\mu}$. We take for $p^{\mu}$
one of the other momenta occurring in the problem. In this
calculation we choose for $p^{\mu}$ the 4-moment of the incident
electron ($p_1^{\mu}$).

For simplicity in the expressions and in the numerical calculation
we assign a number for each 4-moment as it is shown in Fig.1. In
this conditions we represent the products $s(p_i,p_j)$ and
$t(p_i,p_j)$ with the symbols $s_{ij}$ and $t_{ij}$ respectively.
For the auxiliary moments $r_2^1,r_2^2,r_3^1,r_3^2$ we assign the
numbers $4,5,6,7$ respectively. Using of the above rules and
definitions we can write the $T$ amplitudes as follow:

\begin{eqnarray}
T_1(+)&=& 0  ,\nonumber\\
T_1(-)&=&8t_{12}s_{12}t_{24}s_{57}t_{63},\nonumber\\ T_2(+)&=& 8
t_{31}s_{23}t_{36}s_{75}t_{41},\nonumber\\ T_2(-)&=&8
t_{32}(s_{13}t_{36}-s_{12}t_{26}) s_{75}t_{41},\nonumber\\
T_3(+)&=& 2 t_{34}(4 s_{57}t_{61}(s_{26}t_{61}+s_{27}t_{71})-
     2 s_{27}t_{61}(s_{56}t_{61}+
     s_{57}t_{71}+s_{52}t_{21})+ \\
 &&    2 t_{61}s_{27}
     (s_{56}t_{61}+s_{57}t_{71}
     -s_{52}t_{21}) ) ,\nonumber\\
T_3(-)&=& 2 t_{34}(4 s_{57}t_{61}(t_{26}s_{61}+
     t_{27}s_{71}) -
     2 t_{26}s_{71}(s_{56}t_{61}+
     s_{57}t_{71}+s_{52}t_{21})+
     4 s_{51}t_{21}t_{62}s_{27}  + \nonumber\\
&&     2 t_{62}s_{17}
     (s_{56}t_{61}+s_{57}t_{71}-
     s_{52}t_{21}) ) ,\nonumber\\
T_4(+)&=& 2 t_{41} (4 t_{36}s_{75}(s_{26}t_{61}+
     s_{27}t_{71})   -
     2 s_{27}t_{61}(t_{36}s_{65}+
     t_{37}s_{75}+t_{32}s_{25}) +
    4 t_{31}s_{25}t_{62}s_{27}   + \nonumber\\
&&     2 t_{61}s_{27}
     (t_{36}s_{65}+t_{37}s_{75}-
     t_{32}s_{25}) ) , \nonumber\\
T_4(-)&=& 2 t_{41} (4 t_{36}s_{75}(t_{26}s_{61}+
       t_{27}s_{71}) -
     2 t_{26}s_{71}(t_{36}s_{65}+
     t_{37}s_{75}+t_{32}s_{25})    +
     4 t_{32}s_{15}t_{62}s_{27}    + \nonumber\\
&&     2 t_{62}s_{17}
     (t_{36}s_{65}+t_{37}s_{75}-
     t_{32}s_{25}) ) , \nonumber
\end{eqnarray}

After the evaluation of the amplitudes of the corresponding
diagrams,
 we obtain
the cross-sections of the analyzed processes for each point of the
phase space. For the numerical calculation we use a Monte Carlo
computer program, which makes use of the subroutine RAMBO (Random
Momenta Beautifully Organized)\cite{rambo}.

We use the Breit-Wigner propagator for the Majorana neutrino $N$
for different values of the mass $M_N$. The total width $\Gamma_{N
\rightarrow all}$ of $N$ was determined at tree level considering
the dominant decay modes $N\rightarrow W^{\pm}l_j^{\mp}$:
\begin{equation}
\Gamma_{N \rightarrow all}=\frac{g^2 H}{(32 \pi M_N^3 M_W^2)}
(M_N^2-M_W^2)(M_N^4+M_N^2 M_W^2-2 M_W^4)
\end{equation}
In the next section we present our results showing the cross
section for different masses and different center of mass
energies. Moreover we present angular distributions of the final
lepton as a function of the angle with the beam for different
initial photon polarizations.

\section{\bf Results}

Using the helicity formalism we have very compact expressions for
the amplitudes (equations 2,3 and 9).   In $\mid \bar M \mid^2$ we
average over the initial polarization of the electron and sum over
the final polarization of the $W$ and $l^+_j$ and over the flavor
of the final lepton. Moreover an $\frac12$ factor is included to
avoid double counting of the two $W$ when integrating over the
phase space. For the unpolarized cross-section we also have to
average over the initial photon polarization. We take as inputs
the values of $\sqrt{s}$, $M_N$ and $H_1$. The cross section is
formally $\propto H_1$. The $H$ dependence is most complicate due
to the Majorana neutrino propagator . In the
$M_W<M_N<\sqrt{s}-M_W$ kinematic region (Reg.I), where the
intermediate Majorana neutrino may be on-shell, the total
cross-section is almost independent of the $H$ value. In the other
hand in the $M_N>\sqrt{s}-M_W$ region (Reg.II), where the Majorana
neutrino is off-shell, the total cross-section is approximately
proportional to $H$. The behaviour in Reg.I is easy to realize if
we make the so-called peaking approximation, in which the
Breit-Wigner shape of the Majorana neutrino propagator is replaced
by a delta function. In this region the $H$ dependence in the
numerator is canceled by the $H$ factor in  the total width.
Considering only the relevant factors in the cross section, we
have

\begin{eqnarray}
\sigma=\sum_j \sigma_j \sim \cdots H_1 H
\frac{1}{(q^2-M_N^2)^2+M_N^2\Gamma_N^2}\cdots
\end{eqnarray}

where $j$ labels the final lepton flavors. Making now the peaking
approximation

\begin{eqnarray}
\cdots \frac{1}{(q^2-M_N^2)^2+M_N^2\Gamma_N^2}\cdots \rightarrow
\frac{\pi}{M_N \Gamma_N} \delta(q^2-M_N^2)
\end{eqnarray}

and since that $\Gamma \sim H$ (eq.10) then we can see that
$\sigma=\sum_j \sigma_j$ is almost independent of $H$ in Reg.I.

The Fig.2 show the $M_N$ dependence of the unpolarized
cross-section $\sigma/H_1$ at fixed $\sqrt{s}$ for $H=0.1$. We
include the 2-body process ($\gamma e^- \rightarrow W^- N $) to
check the correctness of our final 3-body calculation. In Fig.3 we
show the $\sqrt{s}$ dependence of $\sigma/H_1$ for different
values of $M_N$ keeping again $H=0.1$.

With the helicity formalism that we have used in this calculation
is easy to study distributions of the final lepton for different
polarizations of the initial photon. As an illustration we present
in Fig.4 the angular distribution of the final lepton for left and
right photons and for different values of $M_N$. We have ignored
the experimental difficulties of detecting the discussed process
unambiguously but this kind of distributions could be useful to
reject possible background and for to test no-standard coupling of
this neutrinos \cite{2beta}.

In different classes of models \cite{cotas} $H_1$ and $H$ are
severely restricted by available experimental data (LEP and
low-energy data). This bounds are $H_1<0.016$ and $H<0.122$. In
this work we have used the value $H=0.1$ which agrees whith the
bound over $H$.

To illustrate the possible impact of this process in the discovery
of Majorana neutrinos we show in Fig 5 and Fig 6 the curves with
constant events number in the plane ($H_1,M_N$) for $\sqrt{s}=300$
GeV and $\sqrt{s}=500$ GeV respectively. In both figures we take
$H=0.1$ and we include an upper bound for $H_1$ ($H_1<0.01$). We
have considered a most restrictive value for $H_1$ and $H$ that
the inferred of the experimental bounds such that the considered
upper bound is sufficiently restricted to make a conservative
analysis of the ability of this collider to discover the nature of
the heavy neutrinos. We have used the estimated luminosity
\cite{tesla} for the $\gamma e^-$ collider of ${\cal L} = 100
fb^{-1}$. If we take as reasonable the threshold of 100 events
then we could see signatures for Majorana neutrinos for masses
lower than 250 GeV and 400 GeV for $\sqrt{s}=300$ GeV and
$\sqrt{s}=500$ GeV respectively.

Summarizing, we calculate the cross-section for the process
$\gamma e^- \rightarrow W^- W^- l_j^+$ where $l^+_j$ are light
anti-leptons $(e^+,\mu^+.\tau^+)$. We have included all the
contributions considering that the intermediate Majorana neutrinos
can be either on-shell or off-shell. We study the total
unpolarized cross-section and the angular distribution of the
final lepton for polarized initial photon. Finally we investigate
the events number as a function of $H_1$ and $M_N$ for $H=0.1$ and
for $\sqrt{s}=$ 300 and 500 GeV. We find an important range of
$M_N$ for which would be possible to see signatures for Majorana
neutrinos.
%%%%%%%%%%%%%%%%%%%%%%%%%%%%%%%%%%%%%%%%%%%%%%%%%%%%%%%%%%%%%%%%%%%%%%%%%

{\bf Acknowledgements}

We thank CONICET (Argentina), Universidad Nacional de Mar del
Plata (Argentina) and  CONACyT (Mexico) for their financial
supports.

\pagebreak

\noindent{\large \bf Figure Captions}\\

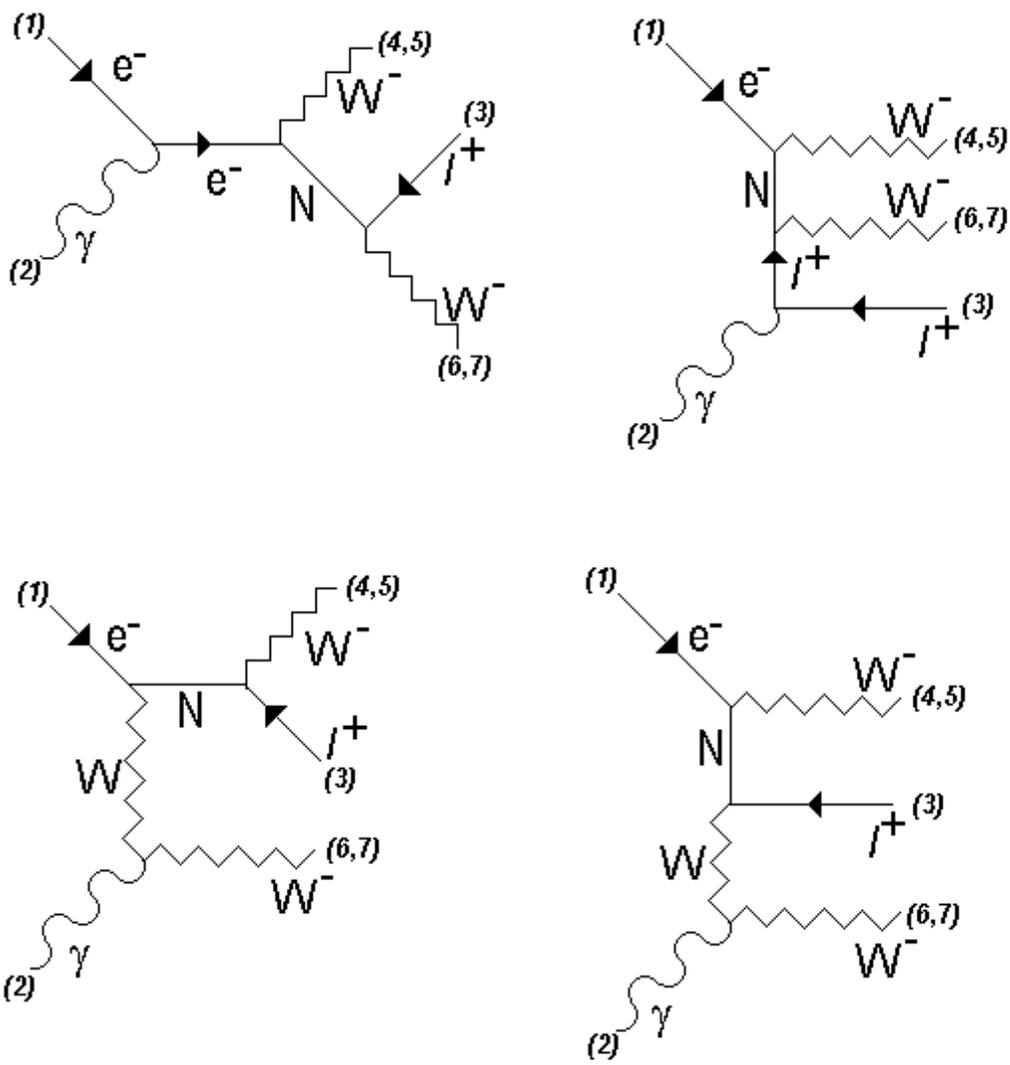
\noindent{\bf Figure 1:} Feynman graph contributing to the
amplitude of the $\gamma e^- \rightarrow W^-W^- l^+$ process.

\noindent{\bf Figure 2:}Unpolarized cross-section as a function of
the Majorana neutrino masses for different center of mass energies
(200, 300 and 500 GeV). The dot-solid line represent the 2-body
process for the same center of mass energies.

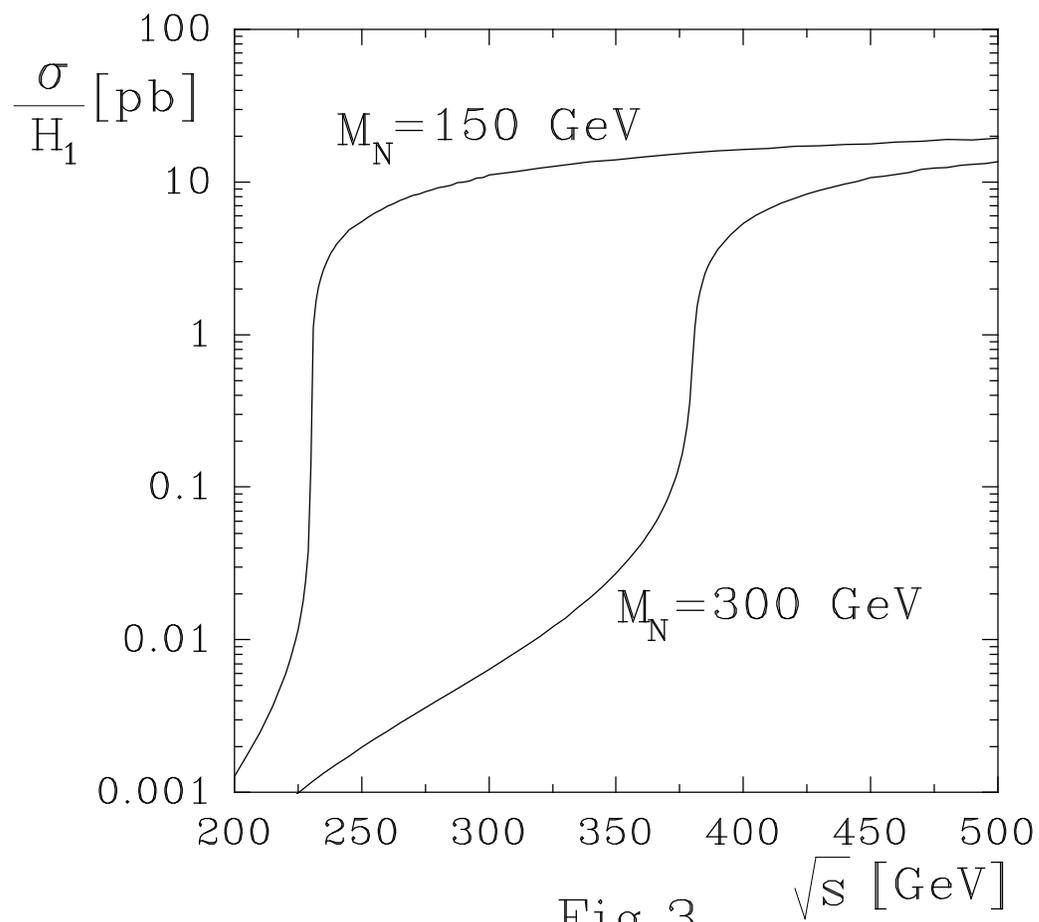
\noindent{\bf Figure 3:} Unpolarized cross section as a function
of the center of mass energies for different Majorana neutrino
masses (150 and 300 GeV).

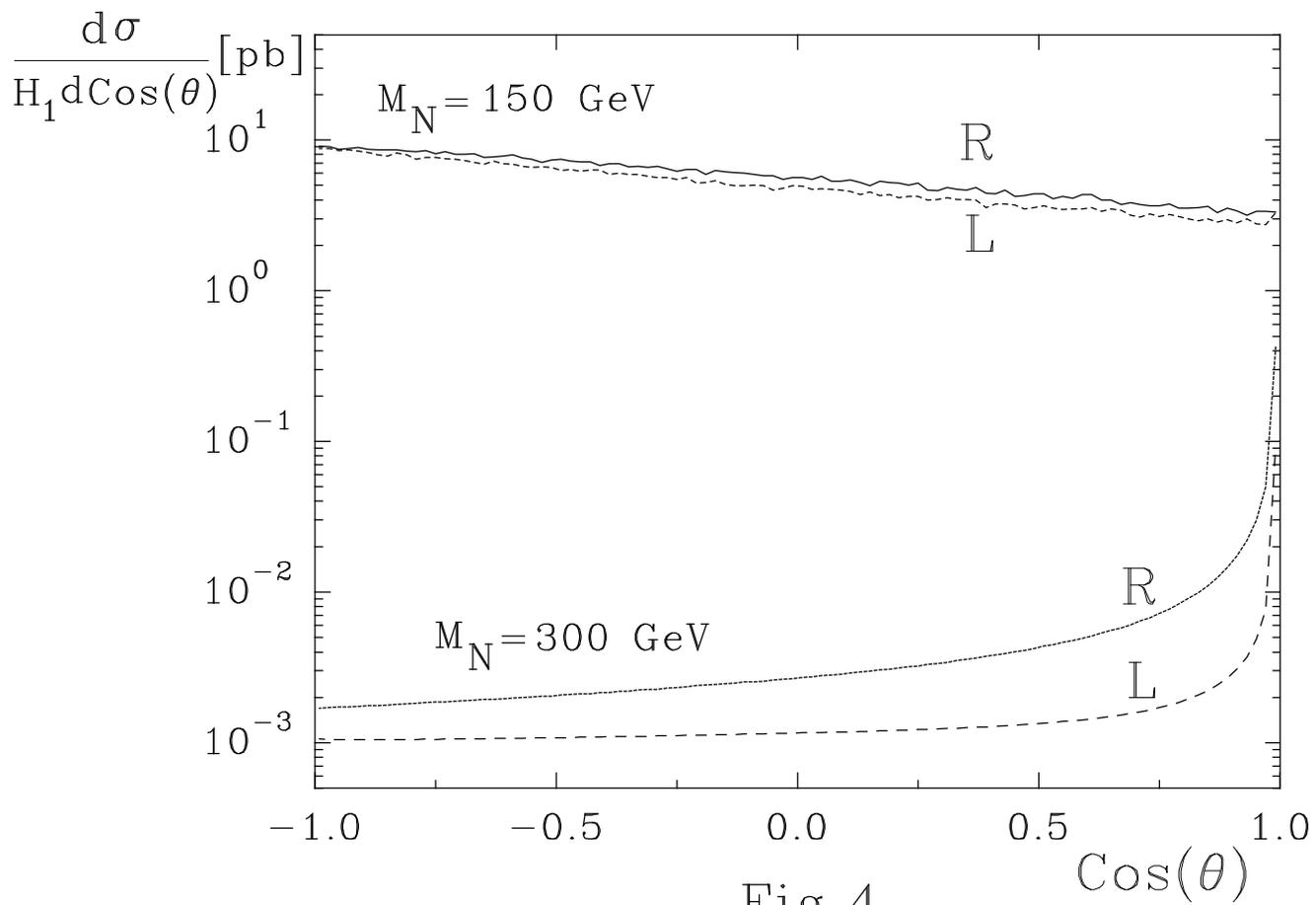
\noindent{\bf Figure 4:} Angular distribution of the final lepton
with the beam axis for polarized initial photon (R: right-handed,
L: left-handed), for $\sqrt{s}=$300 GeV and for two Majorana
neutrino masses (150 and 300 GeV).

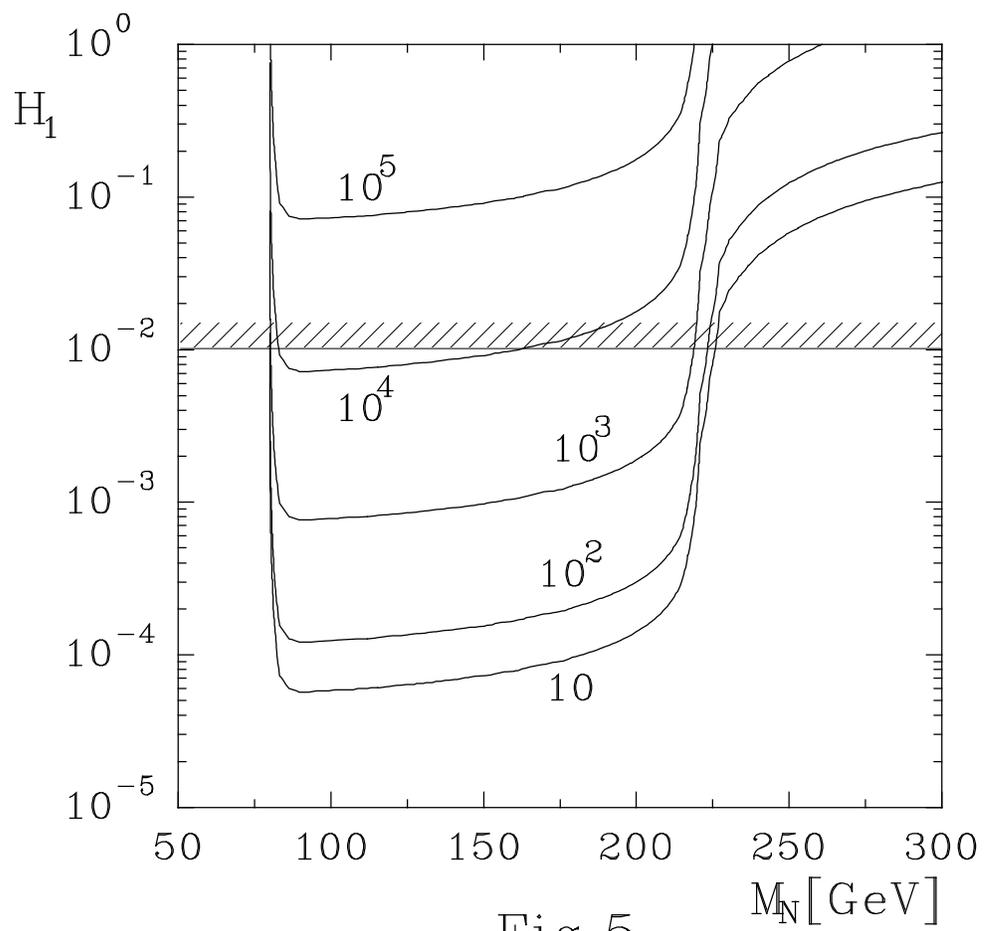
\noindent{\bf Figure 5:} Curves with constant events number($10$,
$10^2$, $10^3$, $10^4$, $10^5$) in the ($H_1$,$M_N$) plane for
$\sqrt{s}=$300 GeV. The dashes line represent an upper bound for
$H_1$.

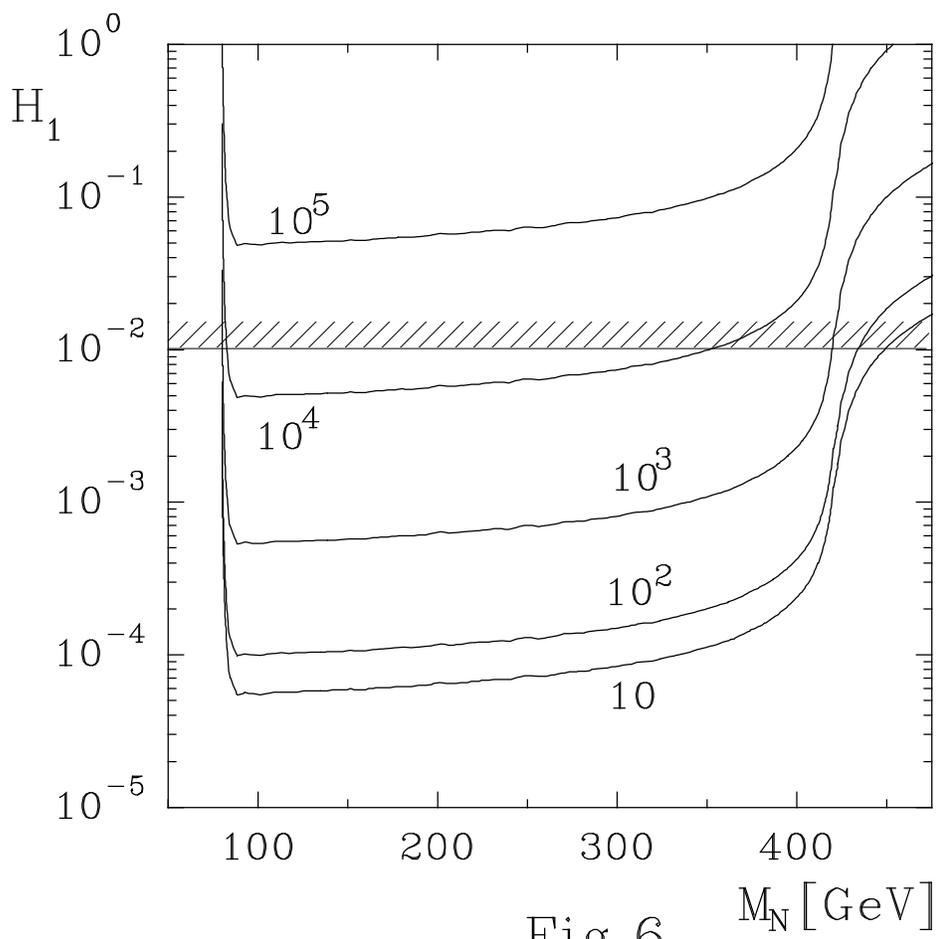
\noindent{\bf Figure 6:} The same of Fig.5 but for $\sqrt{s}=$500
GeV.

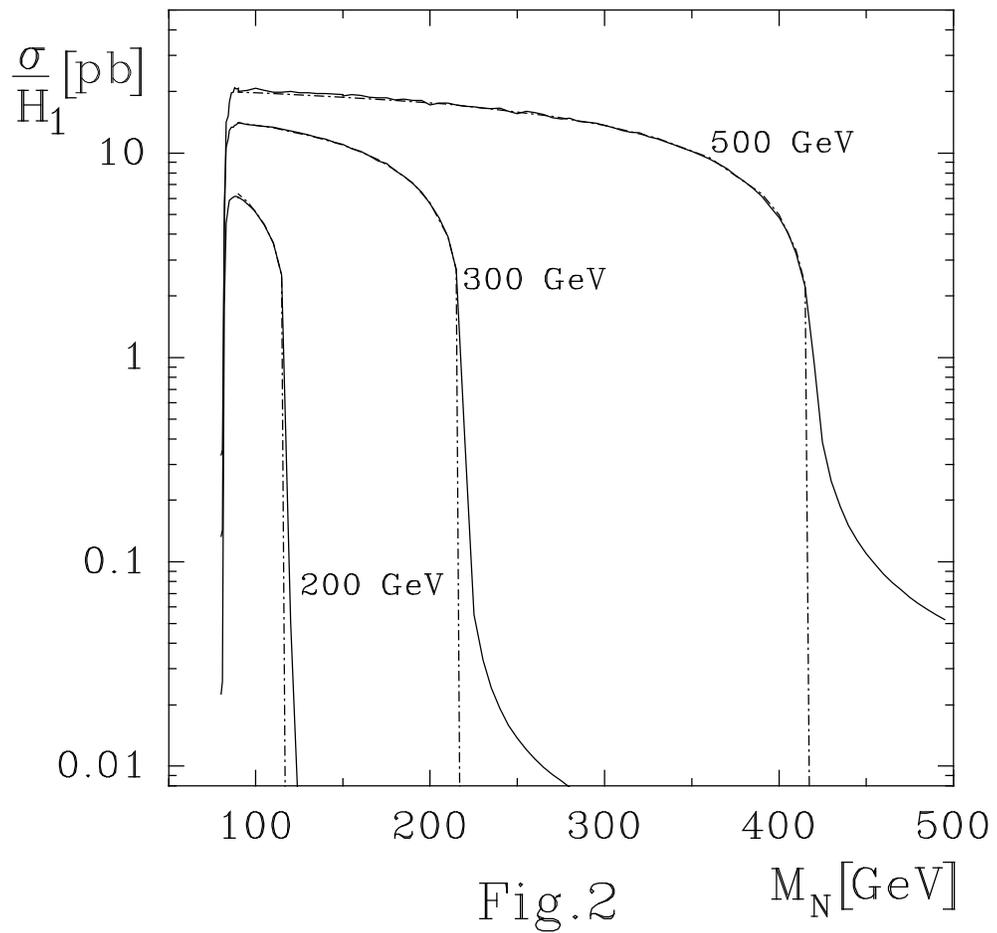
\end{document}